\documentclass[universe,review,accept,pdftex,oneauthor]{Definitions/mdpi} 
\firstpage{1} 
\pubvolume{11}
\issuenum{9}
\articlenumber{304}
\pubyear{2025}
\copyrightyear{2025}
\externaleditor{Laszlo Szabados and Tamas Szalai} % More than 1 editor, please add `` and '' before the last editor name
\datereceived{12 July 2025} 
\daterevised{31 August 2025} % Comment out if no revised date
\dateaccepted{2 September 2025} 
\datepublished{8 September 2025 } 
\hreflink{https://doi.org/\linebreak 10.3390/universe11090304} % If needed use \linebreak
\Title{The Variable Sky Through the OGLE Eye}

\TitleCitation{The Variable Sky Through the OGLE Eye}

\Author{{Patryk}
 Iwanek \orcidA{}}

\AuthorNames{Patryk Iwanek}

\isAPAStyle{%
       \AuthorCitation{Iwanek, P.}
         }{%
        \isChicagoStyle{%
        \AuthorCitation{Iwanek, P.}
        }{
        \AuthorCitation{Iwanek, P.}
        }
}

\address[1]{%
Astronomical Observatory, University of Warsaw, Al. Ujazdowskie 4, 00-478 Warsaw, Poland; piwanek@astrouw.edu.pl}

\abstract{The Optical Gravitational Lensing Experiment (OGLE) is one of the most productive and influential photometric sky surveys in the history of observational astronomy. Originally designed to detect dark matter through gravitational microlensing events, OGLE has evolved into a cornerstone of time-domain astrophysics, delivering three decades of two-band, high-cadence observations \textcolor{black}{of approximately two billion stars} across the Galactic bulge, disk, and Magellanic System. This review summarizes OGLE's key contributions to variable star research, including the discovery, classification and characterization of pulsating stars, eclipsing, ellipsoidal, and rotating variables, or irregular and eruptive stars. Particular emphasis is placed on the OGLE Collection of Variable Stars (OCVS), a publicly available and systematically expanded dataset that has become a fundamental resource for studies of stellar variability and evolution, Milky Way and other galaxies structure, microlensing, compact objects, exoplanets and more. The synergy between OGLE and other major sky surveys, including \textcolor{black}{ASAS, ASAS-SN, ATLAS, Gaia, KMTNet, MACHO, MOA, TESS, PLATO,} or ZTF further amplifies its scientific reach.}


\begin{document}

%%%%%%%%%%%%%%%%%%%%%%%%%%%%%%%%%%%%%%%%%%
\section{Introduction}

Variable stars are not just curiosities---they are a fundamental tool in astronomy. Their light curves convey a lot of knowledge about stellar interiors, binarity, mass loss, rotation, magnetic activity, outbursts, evolution, or~even unseen companions. They serve as distance indicators, tracers of \textcolor{black}{galactic structures}, and~tools for identifying exoplanets or compact objects through gravitational microlensing. \textcolor{black}{Beyond these roles, variable stars also inform about galactic chemical evolution, e.g.,~classical Cepheids map the disk's metallicity gradient, while RR Lyrae stars trace the old, metal-poor halo and accreted substructures---an approach often termed \textit{{galactic archaeology}
}.} This wide applicability has made variable stars indispensable across many fields, from~asteroseismology to cosmology. \textcolor{black}{Looking ahead, asteroseismology will be a key method for determining exoplanet host-star properties with the \textit{PLAnetary Transits and Oscillations of stars} (PLATO) mission, which will deliver high-cadence space-based {photometry} \citep{rauer2025}.} However, modern studies usually require long-term, high-cadence and precise photometric monitoring, which only became widely feasible with the rise of large-scale time-domain~surveys.

Although the formal beginning of variable star astronomy is associated with the discovery of Mira ($o$ Ceti) in 1596, the~variable sky has been observed for centuries. Historical records from East Asia and Europe described transient stellar phenomena, including so-called ``guest stars'', i.e.,~novae and supernovae, which have been confirmed as real astrophysical events. \textcolor{black}{There are also suggestions that the periodic variability of Algol may have been recognized in antiquity---for example by the ancient Egyptians---with its period recorded in the Cairo Calendar for religious reasons \citep{jetsu2013}. As~argued by \mbox{\citet{jetsu2013}}, the~Cairo Calendar may be the oldest preserved historical document attesting to the discovery of a variable star.}

What distinguishes Mira from \textcolor{black}{transient} events is its periodic behavior. In~1596 David Fabricius noticed a new bright, reddish star in the constellation Cetus. He described it as slightly brighter than $\alpha$ Arietis and resembling Mars in color. Fabricius observed the star for several weeks before it faded from view, and~he re-identified it in 1609, albeit without initially recognizing it as the same object. The~periodic nature of Mira's variability was identified in 1638 by Johannes Holwarda. He estimated its period to be about 11~months. After~that, in~1662, Johannes Hevelius named the star \textit{Mira} (“the wonderful one”) and published a dedicated treatise ({\it Historiola Mirae Stellae}), affirming its significance \citep{hevelius1662}. Recent analyses combining historical records with modern photometry confirm that Mira's pulsation period remained remarkably stable over more than four centuries, at~approximately 330 days \citep{2024AN....34530131N}.

A major turning point came in 1784, when British amateur astronomer Edward Pigott discovered the first classical Cepheid---$\eta$ Antinoi (now $\eta$ Aquilae) and measured its period to be approximately $7.18$ days---remarkably close to the modern value of $7.176641$ days. In~the same year, John Goodricke \textcolor{black}{(also British amateur astronomer)} independently discovered the variability of $\delta$ Cephei. He determined the variability period on $5.37$ days (modern measurement is $5.366249$ days). Pigott, the~more experienced observer, acted as a mentor to Goodricke, and~the two friends collaborated closely while living in York. Despite Pigott's priority in identifying the first classical Cepheid, it was Goodricke's work on $\delta$ Cephei that ultimately led to this class of stars being named after the latter object---hence {\it Cepheids} and not {\it Aquilaes} \citep{hoskin1979, bono2024}.

Over a century later, the~study of Classical Cepheids led to one of the most important breakthroughs in cosmology. While working at the Harvard College Observatory, \textcolor{black}{Henrietta Swan Leavitt} investigated variable stars in the Magellanic Clouds. In~1907, she noted a~striking correlation between variability periods and apparent brightness of a subset of variables of the Small Magellanic Cloud \citep{leavitt1907}. In~1912, \textcolor{black}{Henrietta Swan Leavitt} published a refined version of this relation based on 25 variables (classical Cepheids), laying the foundation for the so-called period–luminosity relation (PLR), also known as the {\it Leavitt Law} \citep{leavitt1912}. Despite using only a handful of photographic plates and a provisional magnitude scale, Leavitt achieved remarkable accuracy in estimating both periods and brightness. \textcolor{black}{However, her original relation lacked an absolute zero-point because the distance to the Small Magellanic Clouds was not known at the time. The~first calibration attempts were made by Ejnar Hertzsprung \citep{hertzsprung1907}, and~a more robust zero-point was later provided by Harlow Shapley using distances to globular clusters \citep{shapley1918}. These efforts, along with the subsequent separation of different populations of Cepheids in the following decades (see the historical summary in \citet{fernie1969}), eventually enabled Edwin Hubble to use the PLR for extragalactic distance measurements, leading to the discovery of the expanding Universe \citep{hubble1929}}. A~recent reanalysis of Leavitt's original sample using modern data confirms the overall validity of the relation and shows that most of the measured periods agree with current values to better than $0.01$ days \citep{breuval2025}. The~PLR remains at the core of the cosmic distance ladder and plays a key role in the debate over the value of the Hubble constant \citep{freedman2001, riess2022}.

During the twentieth century, advances in spectroscopy and photometric observations and theoretical modeling enabled the discovery and characterization of numerous other classes of variable stars. Among~pulsators, there were known RR Lyrae stars {(e.g.,}~\cite{vanGent1932, vanGent1933}), $\delta$~Scuti stars {(e.g.,}~\cite{campbell1900, eggen1956}), and~$\beta$ Cephei variables {(e.g.,}~\cite{frost1906, lesh1978, moskalik1992, sterken1993}). Pulsating white dwarfs, such as ZZ~Ceti were identified in the late 1960s and confirmed as a distinct class by the 1970s \citep{landolt1968, mcgraw1979}. In~parallel, other types of variable stars have been discovered, classified, and~analyzed. Eclipsing binaries, such as Algol-type and $\beta$ Lyrae systems, allowed the determination of stellar masses and radii when light curve modeling was combined with spectroscopic data {(see references in} \cite{kopal1959}). Rotational variables were identified through quasi-periodic modulations, and~their variability was first linked to magnetic phenomena (e.g., spots) in chemically peculiar stars like those of $\alpha^2$ Canum Venaticorum type \mbox{{(e.g.,}~\cite{abt1962, peterson1970, preston1974, hensberge1981}).} Eruptive and cataclysmic variables, such as novae or dwarf novae, attracted attention due to their dramatic outburst {(e.g.,}~\cite{vogt1974, warner1975, udalski1988a, udalski1988b, udalski1989, udalski1990, udalski1991, warner1995}). 
These variables offer a unique opportunity to study accretion discs in detail and thus shed light on accretion processes in \mbox{X-ray} binaries, black holes, and~active galactic nuclei \citep{warner1995book}. Moreover, irregular variability was increasingly recognized in young stars of T-Tauri type \citep{appenzeller1989}. All these discoveries broadened the landscape of variable stars taxonomy and revealed the ubiquity of stellar variability across all stages of stellar~evolution.

The characterization and discovery of the variable stars were accelerated by the development of large-scale CCD sky surveys, most notably and pioneering the Optical Gravitational Lensing Experiment (OGLE), launched in the early 1990s \citep{udalski1992}. Originally designed to search for dark matter using microlensing events, OGLE quickly became one of the most prolific and influential photometric sky surveys in astronomy. The~OGLE project provides high-quality, long-term, two-band photometry \textcolor{black}{for approximately two billion stars} in the Large and Small Magellanic Clouds, the~Magellanic Bridge, the~Galactic bulge, and~the Galactic disk. Similar surveys, such as Exp\'erience pour la Recherche d'Objets Sombres (EROS) \citep[][]{aubourg1993}, and~MAssive Compact Halo Objects (MACHO) \citep{alcock1993}, shared this initial focus on microlensing, but~all of them revolutionized time-domain astrophysics through a by-product of their observational strategy: the continuous monitoring of the~sky.

The operational design of the OGLE, EROS, and~MACHO sky surveys led to the discovery of a large number of variable stars, increasing the number of known variable stars by orders of magnitude \citep{udalski1994a, udalski1995a, udalski1995b, udalski1996, udalski1997a, alcock1997a, alcock1997b, beaulieu1995, deLaverny1998}. These by-products not only filled previously unexplored regions of the sky but also provided the first large samples of many types of variables in the Milky Way, and~in the extragalactic environments. This legacy transformed the role of the microlensing surveys into a ``manufacture'' of variable star~discoveries.

Due to the sheer breadth of discoveries enabled by the OGLE survey over the past three decades, this review focuses specifically on results from the last few years of the project. This paper is organized as follows. In~Section~\ref{sec:ogle}, I provide an overview of the OGLE project, its objectives, instrumentation, sky coverage, and~survey strategy. Section~\ref{sec:ocvs} introduces the structure and scope of the OGLE Collection of Variable Stars (OCVS). In~Section~\ref{sec:variable_stars}, I~highlight key discoveries across different classes of variable stars. Finally, in~Section~\ref{sec:summary}, I~conclude the~paper.

\section{Overview of the OGLE~Project} \label{sec:ogle}

The OGLE project is one of the longest-lasting and most impactful photometric sky surveys in the history of astronomy. Initiated in 1992, it was originally designed to search for gravitational microlensing events, following Bohdan Paczyński's idea to detect dark matter in the form of massive compact halo objects \citep{paczynski1986}. Since then, OGLE has grown into a versatile, large-scale time-domain survey dedicated to monitoring stellar variability on multiple timescales. Over~the decades, the~project has undergone several major instrumental and operational upgrades, each marking a new phase with increased sensitivity, sky coverage, and~scientific~reach.

\begin{figure}[H]
\centering
\includegraphics[angle=0, scale=0.65]{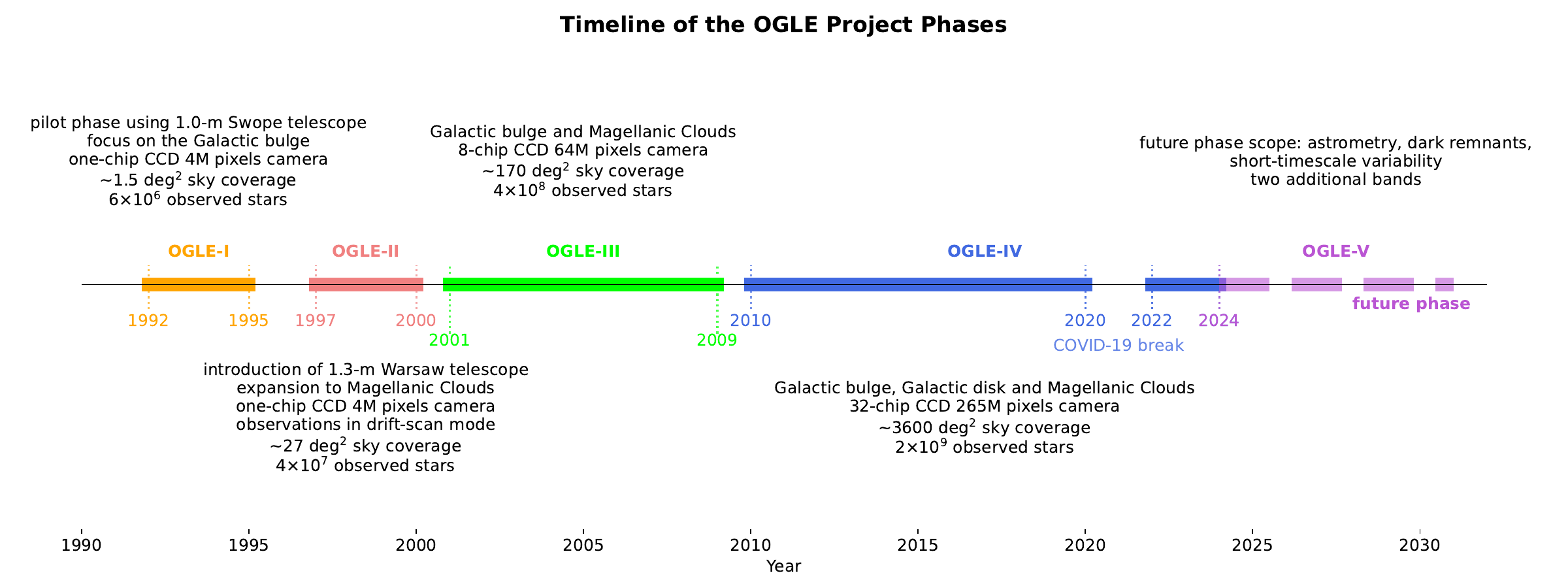}
\caption{{Timeline} of the OGLE project phases with basic characteristics for each~phase.}
\label{fig:timeline}
\end{figure}

\begin{figure}[H]
\includegraphics[scale=0.275]{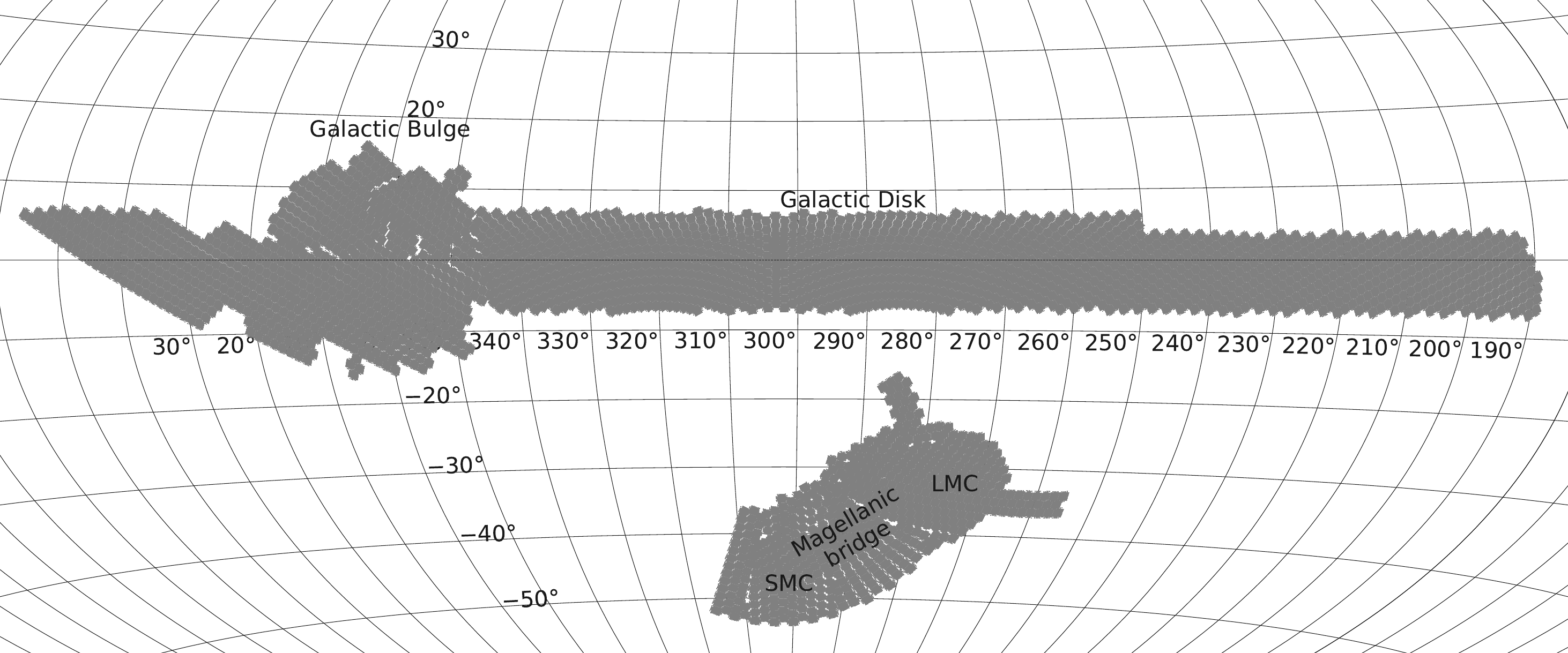}
\caption{{OGLE}-IV sky coverage of the Galactic bulge, the~Galactic disk, and~the Magellanic System, presented in the Galactic~coordinates.}
\label{fig:sky_coverage}
\end{figure}

The OGLE project began in the early 1990s with a pilot phase (OGLE-I) using a 1.0-m Swope telescope at Las Campanas Observatory, Chile, primarily targeting the Galactic bulge \citep{udalski1992}. The~initial effort demonstrated the feasibility of long-term CCD monitoring of dense stellar fields by detecting the first gravitational microlensing events \citep{udalski1993}. In~the second phase (OGLE-II), a~dedicated 1.3-m Warsaw telescope was constructed, significantly enhancing the survey's capabilities and extending coverage to the Magellanic Clouds \citep{udalski1997b}. Later phases, OGLE-III and OGLE-IV, introduced significant instrumental improvements, particularly in the CCD mosaic cameras, enabling growth in both sky coverage and the number of observed stars \citep{udalski2003, udalski2008, udalski2015}. Today, OGLE-IV used a 32-chip CCD camera and has monitored \textcolor{black}{approximately two billion stars} across more than 3600 deg$^2$ of the sky, including the Galactic bulge, Galactic disk, Large and Small Magellanic Clouds, and~Magellanic Bridge. The~observations are conducted in two photometric bands: Johnson {\it V} and Kron-\textcolor{black}{Cousins} {\it I}. The~basic characteristics of each phase of the OGLE project are summarized in Figure~\ref{fig:timeline}. Figure~\ref{fig:sky_coverage} presents the current OGLE-IV sky~footprint.

The most recent phase, OGLE-V, officially launched in 2025, introduces a major conceptual shift in the project scope. Alongside ongoing photometric monitoring, OGLE-V integrates high-precision astrometry as a fundamental component of the survey. One of its primary goals is the identification of microlensing events caused by dark, compact objects such as isolated stellar-mass black holes, which are promising targets for astrometric follow-up studies. Additionally, OGLE-V aims to investigate short-timescale variability (on the order of minutes to hours), opening a new variability window to future discoveries, as~this regime has remained largely unexplored by previous long-term surveys. The~new phase also plans to \textcolor{black}{extend} existing filters with a {\it B}-band and prioritize monitoring of the Magellanic System. As~a result, this will enable three-band {\it BVI} coverage of Large and Small Magellanic Clouds, and~the Magellanic Bridge. OGLE observations will also be carried out in a new wide {\it R}-band (which combines {\it $\mathrm{R_C}$}- and {\it $\mathrm{I_C}$}-bands). The~development of OGLE-V has been supported by a grant awarded in 2025 by the Polish National Science Center to the PI of the project---Andrzej Udalski. Implementation of this new phase is currently~underway.

\startlandscape
\finishlandscape
\unskip

\section{The OGLE Collection of Variable~Stars} \label{sec:ocvs}

The OGLE Collection of Variable Stars (OCVS) is one of the most extensive and diverse databases of stellar variability in modern astrophysics. Compiled over decades of photometric monitoring, OCVS includes more than a million variable stars of different types. It represents a fundamental by-product of the OGLE observational strategy, originally designed to detect microlensing events, but~which also enabled systematic and long-term monitoring \textcolor{black}{of approximately two billion stars.}

The OCVS covers a wide range of astrophysical environments, including the densest stellar fields of the Galactic bulge, the~spiral arms and disk of the Milky Way, as~well as the extragalactic systems of the Magellanic Clouds (as shown in Figure~\ref{fig:sky_coverage}). Each of the major regions---the Galactic bulge (BLG), the~Galactic disk (GD), the~Large Magellanic Cloud (LMC), and~the Small Magellanic Cloud (SMC) has its own dedicated set of variable star catalogs, which are systematically expanded over time. \textcolor{black}{For cluster environments, OGLE produced dedicated samples of variables in star clusters of the Magellanic Clouds enabling cluster-by-cluster occurrence and properties studies ({see e.g.,}~\cite{pietrzynski1998, pietrzynski1999a, pietrzynski1999b}).}

\begin{figure}[H]

\begin{adjustwidth}{-\extralength}{0cm}
\centering
\includegraphics[scale=0.5]{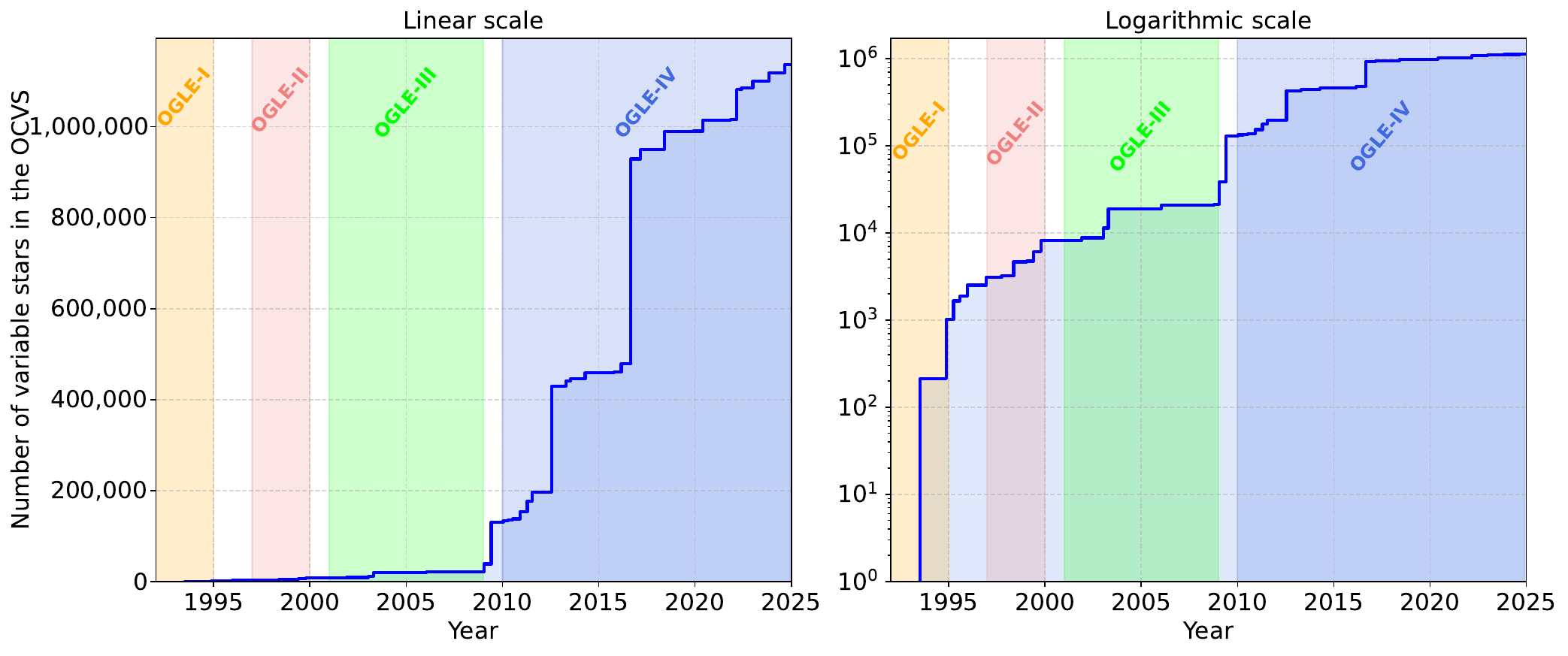}
\end{adjustwidth}
\caption{{Number} of variable stars published by the OGLE project in the years 1992–2025.
{Left} and {right} panels show the number of variable stars in the linear and logarithmic scales,
respectively.}
\label{fig:ocvs_history}
\end{figure}

The collection includes a broad set of variability classes, ranging from pulsating stars to eclipsing and ellipsoidal binaries, rotating and eruptive variables, and~other rare and exotic objects. Table~\ref{tab:ocvs} summarizes the number of variable stars discovered and published in the OCVS. The~collection continues to grow over time, reflecting both the increase in sky coverage and the improved classification techniques. In~Figure~\ref{fig:ocvs_history}, I show how the cumulative number of variable stars in the OCVS has evolved over the years, with~distinctions corresponding to successive OGLE~phases.

\textcolor{black}{Over decades of sky surveys, the~number of known e.g.,~classical pulsators in the Magellanic Clouds has increased dramatically---from only a few thousand at the beginning of the twentieth century to nearly complete samples today. {Figure~3} in \citet{soszynski2018b} clearly illustrates this transformation, showing the steep rise in the cumulative number of classical Cepheids, RR Lyrae stars, type II Cepheids, and~anomalous Cepheids following the OGLE era.}

\textls[-15]{A star in the OCVS is classified as variable when its light curve exhibits statistically significant periodic or non-periodic brightness changes that exceed the typical photometric noise level, with~the detection thresholds depending primarily on mean magnitude, amplitude, and~time-series sampling. For~periodic variables, the~detection efficiency naturally decreases toward low amplitudes and very long periods, while crowding and sky position affect the achievable photometric precision, and~thus the sensitivity to low-amplitude~variability.}

\unskip

\begin{table}[H]
%\begin{center}
\caption{Types and number of variable stars published in the~OCVS.}
\begin{tabularx}\textwidth{lR}
\toprule

\textbf{Type of Variable Stars} & \textbf{Number of Stars} \\ \midrule
Classical Cepheids & {11,689} \\
Type II Cepheids & $2046$ \\
Anomalous Cepheids & $393$ \\
RR Lyrae stars & 129,740 \\
$\delta$ Scuti stars & 42,672 \\
Blue Large-Amplitude Pulsators & $94$ \\
Long-Period Variables & 403,636 \\
Eclipsing binaries & 525,998 \\
Heartbeat stars & $991$ \\
Rotating variables & 18,443 \\
Short-period eclipsing variables & $242$ \\
R Coronae Borealis stars & $23$ \\
Dwarf novae & $1091$ \\
Double Periodic Variables & $32$\\ \midrule
{\bf {Total:}} & \textbf{1,137,090}\\
\bottomrule 
\label{tab:ocvs}
\end{tabularx}
%\end{center}
\end{table}\vspace{-10pt}

The entire collection is publicly available and can be accessed via the OGLE website or directly through the catalog {repository:}

\begin{itemize}
\item	\url{https://ogle.astrouw.edu.pl} $\rightarrow$ {OGLE} Collection of Variable~Stars \newline (accessed on 11 July 2025)
\item {\url{https://www.astrouw.edu.pl/ogle/ogle4/OCVS/}} \newline (accessed on 11 July 2025)
\end{itemize}

\noindent {The catalog} data include observational parameters of the variable stars, such as coordinates, periods, mean magnitudes, brightness amplitudes, color indices, cross-matches with external catalogs, etc. Additionally, time-series OGLE photometry in the {\it I}- and {\it V}-bands is made available to the astronomical community. It is worth emphasizing that the OGLE classification pipeline is still semi-automatic, based on a combination of time-series analysis tools, and~visual inspection of each light curve by team members. This led to exceptionally high completeness and a low contamination~rate.

Due to its scale, homogeneity, rich variability content, and~long-term precise light curves, the~OCVS has also become a perfect dataset for training machine learning models for variability classification and time series analysis. Its structure, completeness, and~diversity of variability types make it an ideal benchmark for testing and validating modern algorithms {(e.g.,}~\cite{richards2012, aguirre2019, godines2019, becker2020, mroz2020a, monsalves2024, li2025}).

\section{The OGLE Contributions to Variable Star~Research} \label{sec:variable_stars}
\unskip

\subsection{Classical~Cepheids}

Classical Cepheids is one of the most important classes of variable stars in modern astrophysics. Their high luminosities and well-defined PLR make them indispensable standard candles for measuring distances within the Milky Way and to nearby~galaxies.

The OGLE project has provided the most complete and homogeneous collection of classical Cepheids in the Milky Way and in the Magellanic Clouds. Most of the classical Cepheids known today in the Magellanic Clouds \citep{soszynski2017a}, and~approximately half of the currently known Galactic Cepheids \citep{pietrukowicz2021} have been identified in the OGLE photometric databases. As~a result, OGLE has essentially completed the task of cataloging classical Cepheids in the Magellanic System---a work that was originally initiated by Henrietta Leavitt at the beginning of the twentieth~century.

The two satellite galaxies---Large and Small Magellanic Cloud---are ideal laboratories for studying classical Cepheids. They are close enough that individual stars are easily resolved with ground-based telescopes, yet distant enough that their stellar populations can be treated as lying at nearly the same distance. The~distance to the Large Magellanic Cloud is measured with remarkable precision---about $1\%$ uncertainty \citep{pietrzynski2019}. This makes the Large Magellanic Cloud, and~by extension the Small Magellanic Cloud, critical anchor points for calibrating the cosmic distance~ladder.

The OGLE Cepheid samples have enabled detailed mapping of the three-dimensional structures of both Clouds. OGLE data show that the Large Magellanic Cloud disk is inclined at $24.2^\circ \pm 0.7^\circ$, with~a position angle of $151.4^\circ \pm 1.7^\circ$, and~exhibits a prominent northern warp \citep{jacyszyn2016}. The~Small Magellanic Cloud, in~contrast, shows a more complex bimodial structure \citep{jacyszyn2016}. Moreover, classical Cepheids have been used to trace the Magellanic Bridge---a structure that links the Large and Small Magellanic Clouds \citep{jacyszyn2020}.

\begin{figure}[H]

\begin{adjustwidth}{-\extralength}{0cm}
\centering
\includegraphics[scale=0.5]{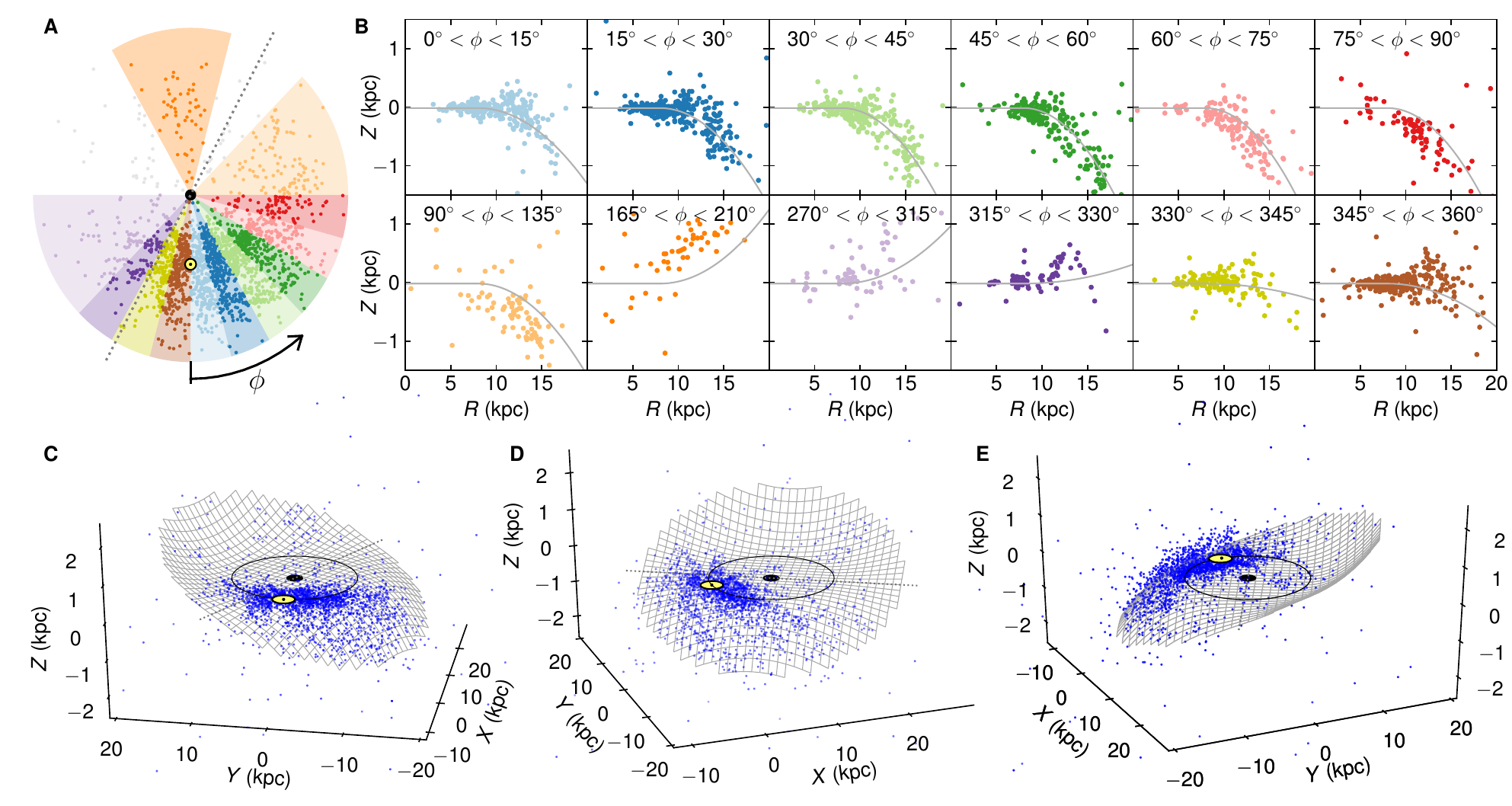}
\end{adjustwidth}
\caption{{Warping} of the Milky Way disk. For~more information see \citet{skowron2019}. \textcolor{black}{Reproduced with The American Association for the Advancement of Science permission.}}
\label{fig:skowron2019}
\end{figure}

The use of OGLE-based Cepheids has revolutionized our view of the three-dimensional structure of the Milky Way. In~a groundbreaking study, the~OGLE team utilized a sample of $2400$ classical Cepheids complemented with Cepheids from other surveys to construct the most detailed map of the Galactic young stellar disk to date \citep{skowron2019}. This study revealed that the Milky Way disk is strongly warped and flared, especially in its outer parts, with a~vertical displacement that reaches up to $\sim$1 kpc from the Galactic plane. The~warping of the Milky Way disk is shown in Figure~\ref{fig:skowron2019}. The~spatial distribution of these young variables also demonstrated clear spiral-arm structures and asymmetries between the nothern and southern parts of the disk. This study provided a direct stellar-based view of the Galaxy's morphology, offering constraints on the mechanisms responsible for disk warping. In a~recent paper \citet{skowron2025} re-estimated distances to the Milky Way Cepheids using mid-infrared photometry and three-dimensional extinction maps. Their results show high consistency with the Gaia parallaxes, with~typical distance uncertainties around $6\%$.

\begin{figure}[H]
\includegraphics[scale=0.85]{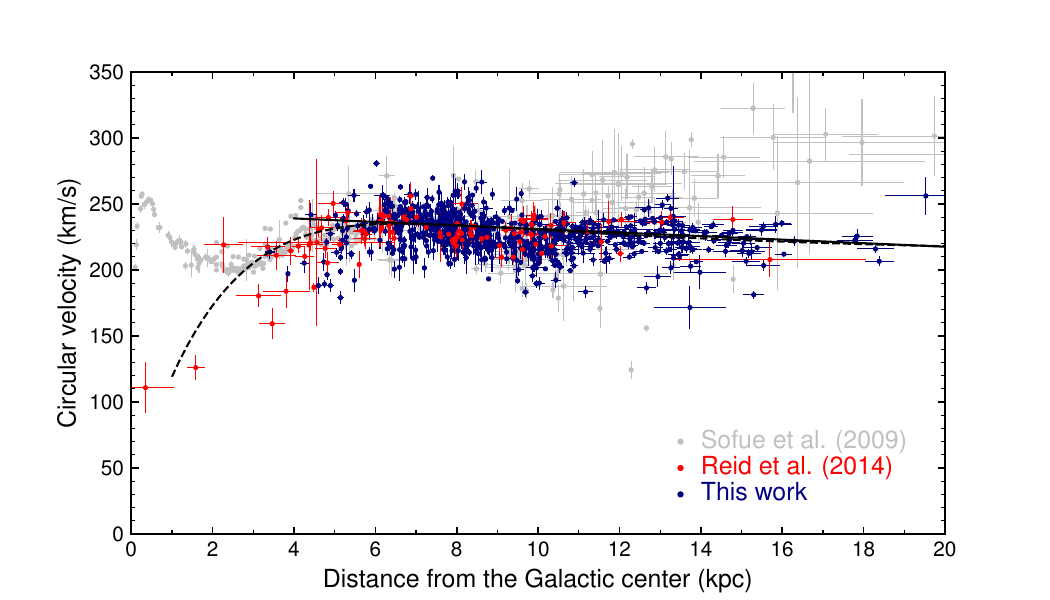}
\caption{{Rotation} curve of the Milky Way obtained from classical Cepheids. \textcolor{black}{The label ``This work'' is part of the original figure and refers to \citet{mroz2019}, not to the present manuscript. For~more information see \citep{sofue2009, reid2014, mroz2019}. © AAS. Reproduced with permission.}}
\label{fig:mroz2019}
\end{figure}

In addition to classical Cepheids' role as distance indicators and tracers of the Galactic structure, they have been used to probe the kinematics of the Milky Way. \citet{mroz2019} combined the OGLE photometry with radial velocities and proper motions from Gaia for 773 Cepheids, to~measure the rotation of the Galactic disk. The~authors measured the rotation velocity as $233.6 \pm 2.8$ km/s. The~rotation curve of the Milky Way is presented in Figure~\ref{fig:mroz2019}.

\unskip

Among the most notable individual discoveries is OGLE-GD-CEP-1884, the~classical Cepheid with the longest known pulsation period in the Milky Way---$78.14$ days---found in the Galactic disk \citep{soszynski2024}. This ultra-long-period Cepheid lies $\sim$4.5 kpc from the Sun and demonstrates that the known population of such rare and massive variables is still~\mbox{incomplete}.

\subsection{Type II and Anomalous~Cepheids}

\begin{figure}[H]

\begin{adjustwidth}{-\extralength}{0cm}
\centering %% If there is a figure in wide page, please release command \centering, for Table, ``\textwidth" should be ``\fulllength"
\includegraphics[scale=0.3]{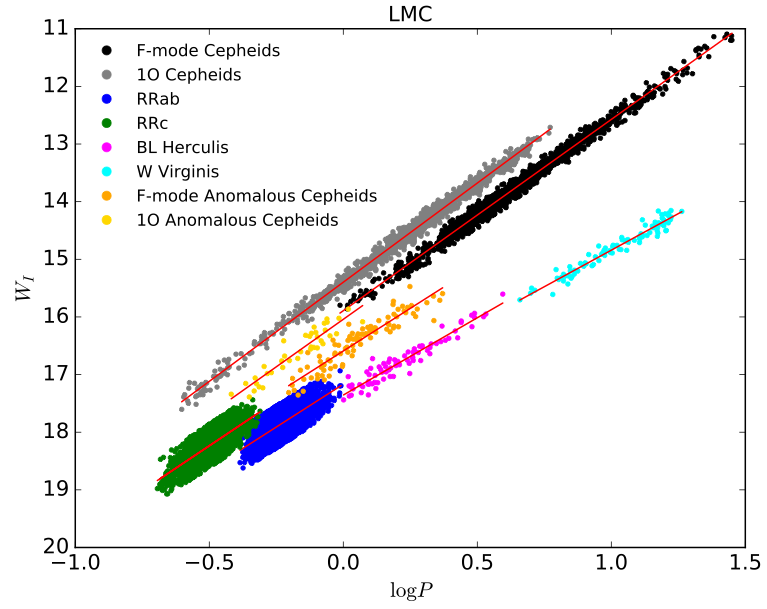} 
\includegraphics[scale=0.3]{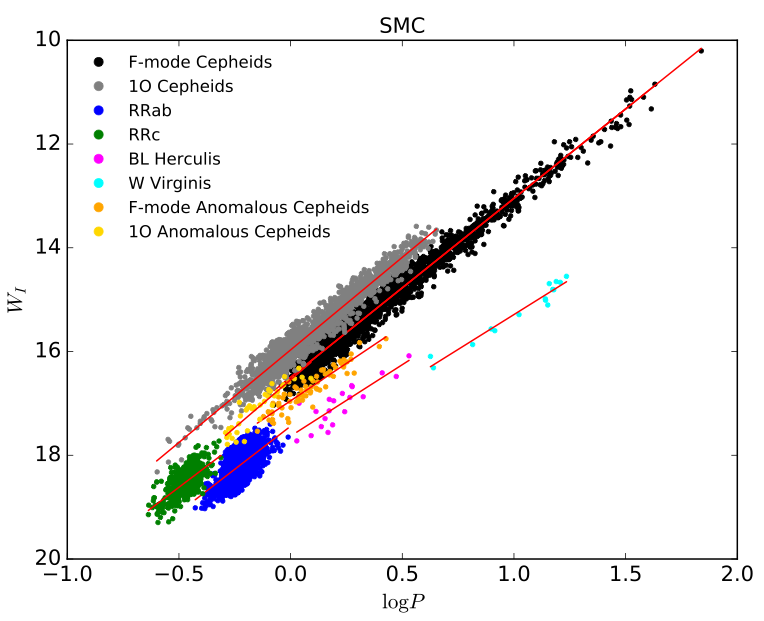}
\end{adjustwidth}
\caption{Period-Wesenheit index relations for different groups of classical pulsators in the Large and Small Magellanic Clouds, including relations for type II and anomalous Cepheids. For~more information see \citet{iwanek2018}.}
\label{fig:iwanek2018_plr}
\end{figure}

Type II and anomalous Cepheids represent less massive and poorly understood counterparts of classical Cepheid pulsators. Although~they also follow PLRs (see Figure~\ref{fig:iwanek2018_plr} with Period-Wesenheit index\endnote{The Wesenheit index is an extinction-free quantity, defined as $W_I = I - 1.55(V-I)$, where {\it I} and {\it V} are apparent mean magnitudes of the stars \textcolor{black}{(see e.g.,~\citet{soszynski2011})}.} {relations,} their evolutionary origins, pulsation periods, and~population characteristics are different. Thanks to the extensive OGLE database, these stars have been discovered and analyzed with unprecedented completeness and~detail.

The OGLE project has provided almost complete collections of type II and anomalous Cepheids to date. In~the Magellanic Clouds, OGLE has discovered 344 type II Cepheids, divided into: 121 BL Herculis, 123 W Virginis, 34 peculiar W Virginis, 66 RV Tauri stars~\citep{soszynski2018a}, along with 271 anomalous Cepheids identified as a separate class \citep{soszynski2015}. The peculiar W~Virginis stars---distinguished by their brighter and bluer appearance, as~well as light curve shapes---emerged as a newly recognized subclass with distinct spatial and evolutionary properties \citep{soszynski2008}. In~the Milky Way, the~OGLE survey contributed to the census of pulsating stars by discovering 1641 type II Cepheids and 119 anomalous Cepheids---the first anomalous Cepheids found in this region \citep{soszynski2017b}.

\textcolor{black}{Particularly} noteworthy is the first-ever detection of type II Cepheids pulsating exclusively in the first overtone mode \citep{soszynski2019}, as~well as a triple-mode anomalous Cepheid~\citep{soszynski2020}.

\begin{figure}[H]
%\begin{center}
\includegraphics[scale=0.45]{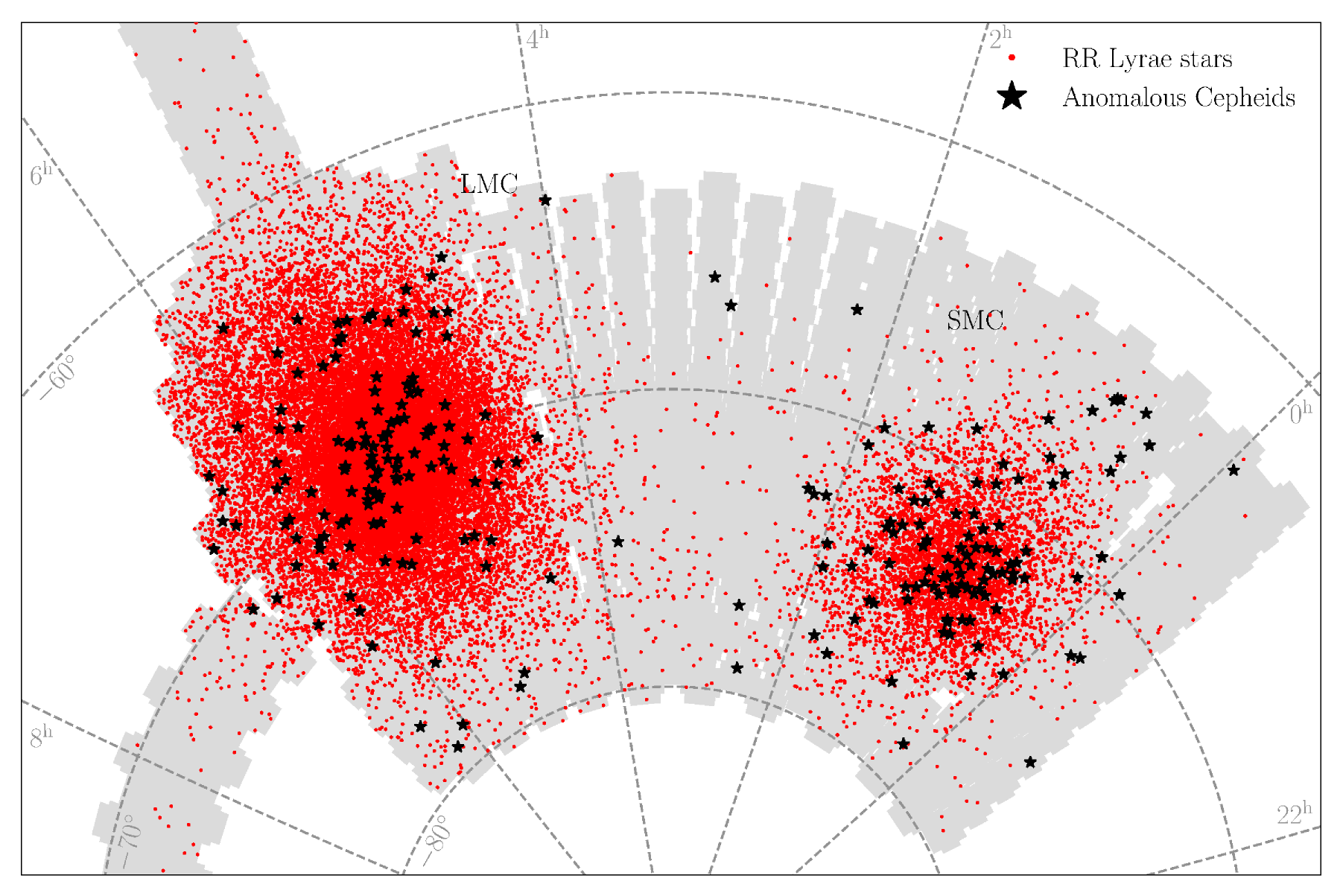} \\
\includegraphics[scale=0.18]{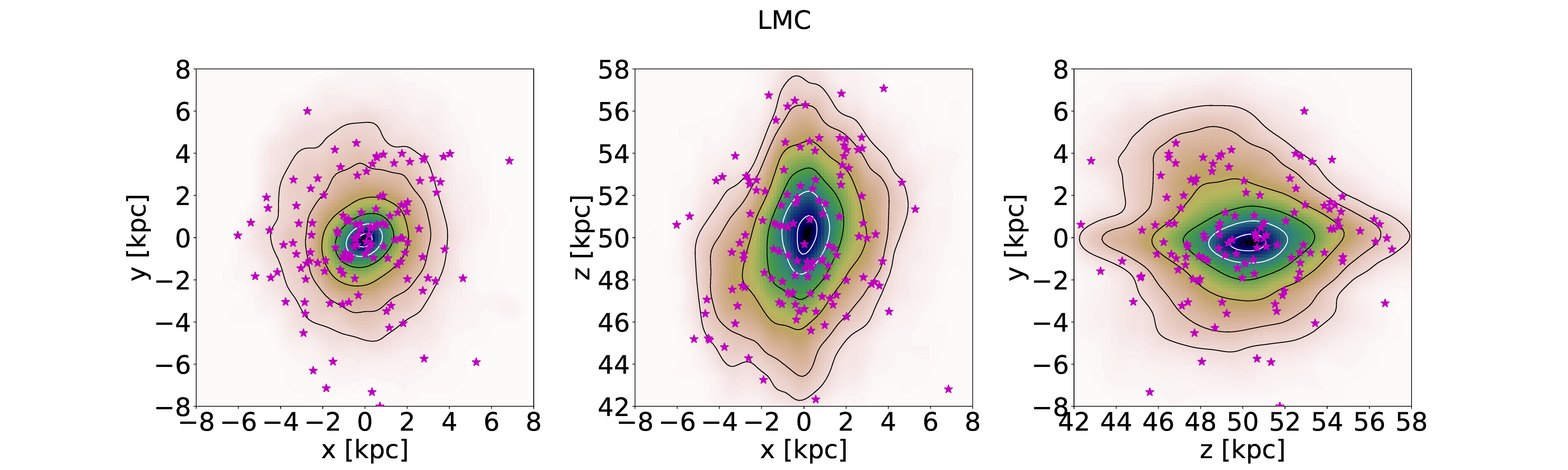} \\
\includegraphics[scale=0.18]{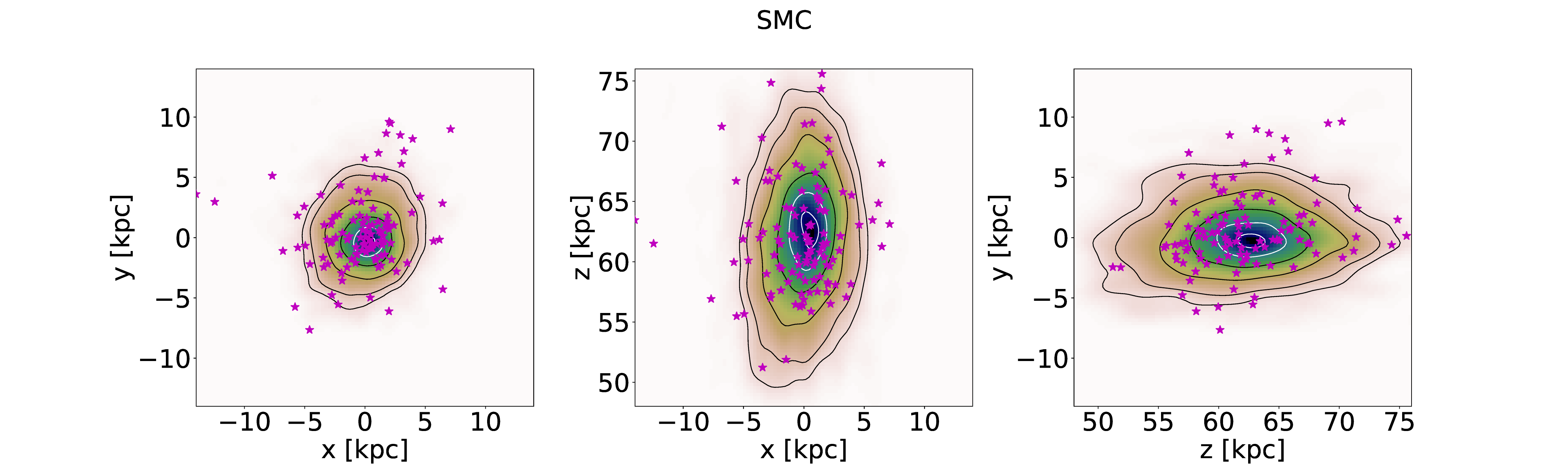} \\
%\end{center}
\caption{{Spatial} distribution of anomalous Cepheids in comparison to the distribution of RR Lyrae stars. Top panel presents equal-area Hammer projection of the MS, while middle and bottom panels present Cartesian projections for the Large and Small Magellanic Cloud, respectively. In~the middle and bottom panels, RR Lyrae stars are shown as a color map with density contours, while anomalous Cepheids are shown as magenta points. For~more information see \citet{iwanek2018}.}
\label{fig:iwanek2018_acep}
\end{figure}

Beyond individual discoveries, OGLE data have enabled population-scale analyses. Using three-dimensional spatial distributions of classical, type II, and~anomalous Cepheids, as~well as the RR Lyrae star in the Large and Small Magellanic Clouds, \citet{iwanek2018} demonstrated that BL Her stars share the old stellar halo with the RR Lyrae variables, while W Virginis stars show characteristics of both old and intermediate age populations, potentially indicating a~mixed evolutionary origin. The~anomalous Cepheids were found to differ significantly from classical Cepheids in spatial distribution, yet show partial overlap with RR Lyrae stars---supporting the binary coalescence scenario as a viable evolutionary channel. The~comparison of spatial distributions of RR Lyrae stars and anomalous Cepheids is shown in Figure~\ref{fig:iwanek2018_acep}.

\subsection{Miras}

Mira variables are long-period, large-amplitude pulsating Asymptotic Giant Branch stars, representing an evolved phase of low- and intermediate-mass stellar evolution. Due to their high luminosities and well-defined PLRs, Miras serve as excellent tracers of old- and intermediate-age stellar populations and can act as valuable distance indicators across a wide range of~wavelengths.

The OGLE project has provided the most complete sample of Mira variables in Milky Way and Magellanic System. This enabled a series of detailed studies exploring their physical, photometric, and~spatial properties. \citet{iwanek2021a} based on Miras from Large Magellanic Cloud, derived precise mid-infrared PLRs. The~authors based their findings on the Spitzer and Wide-Field Infrared Survey Explorer (WISE) observations. Seperate calibrations for oxygen-rich (O-rich) and carbon-rich (C-rich) Miras yielded distances with an accuracy of 5--12\%, confirming that Miras are reliable standard candles in the mid-infrared~domain.

In a subsequent paper, \textcolor{black}{the authors} examined multiwavelength variability from the optical to the mid-infrared, highlighting how pulsation amplitudes systematically decrease with increasing wavelength, while phase lags increase \citep{iwanek2021b}. These effects are especially prominent for C-rich Miras, many of which exhibit dust ejections. The~study also introduced~{synthetic PLRs} for 42 photometric bands, enabling applications for distance estimation in a wide range of surveys, such as OGLE, the~VISTA Near-Infrared YJKs Survey of the Magellanic Clouds System, Legacy Survey of Space and Time,
Gaia, Spitzer, WISE, the~James Webb Space Telescope, the~Nancy Grace Roman Space Telescope (formerly WFIRST), and~the Hubble Space~Telescope.

\begin{figure}[H]
%\begin{center}

\begin{adjustwidth}{-\extralength}{0cm}
\centering
\includegraphics[scale=0.6]{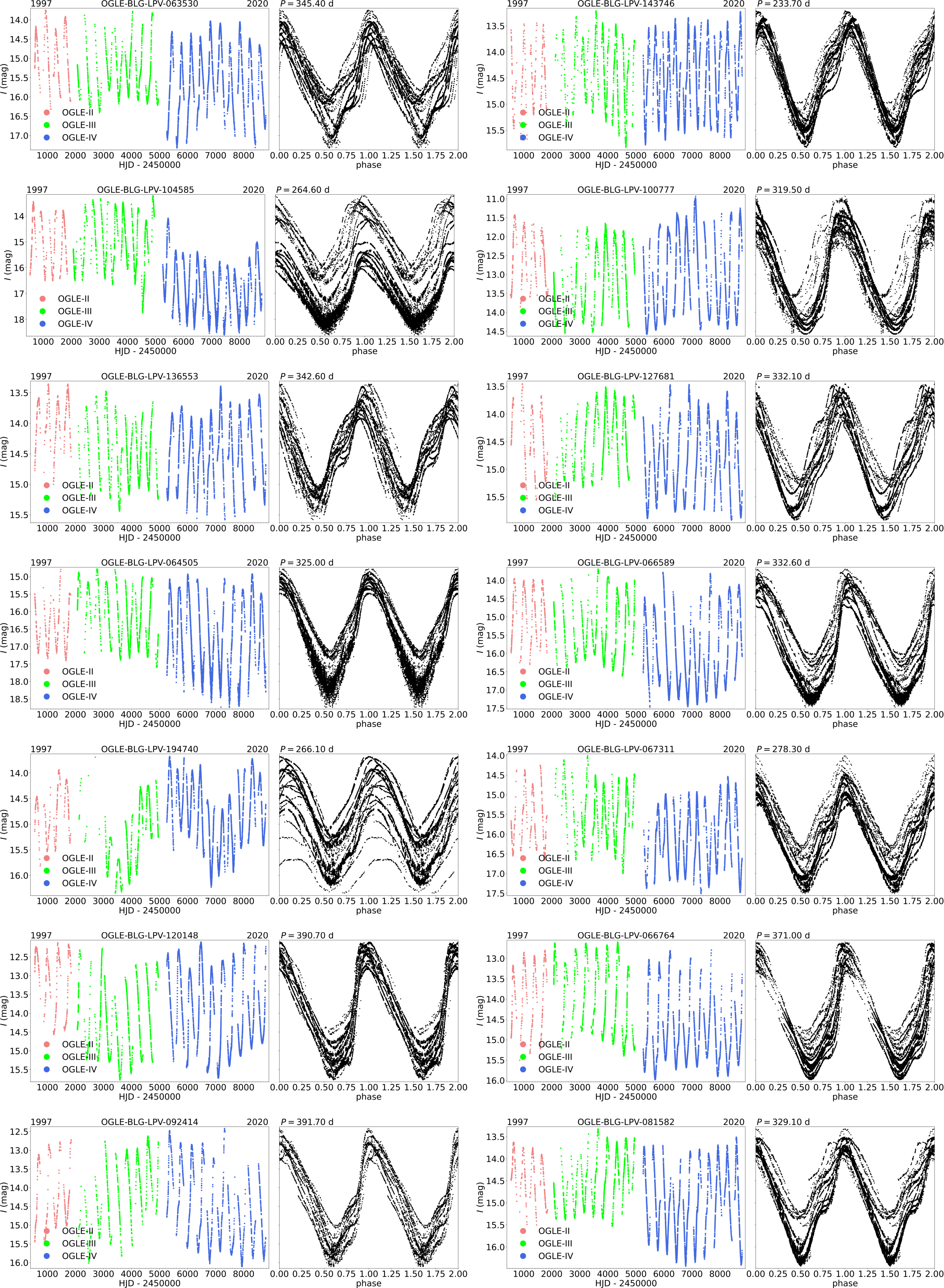}
\end{adjustwidth}
%\end{center}
\caption{{Fourteen} examples of Mira stars from the OCVS which were observed since 1997, i.e.,~the beginning of the OGLE-II phase, until~March 2020. Left-hand-side panels show unfolded light curves, while right-hand-side panels show phase-folded light curves with pulsation periods $P$ (provided above the plots). For~more information see \citet{iwanek2022}.}
\label{fig:iwanek2022}
\end{figure}

In 2022, \citet{iwanek2022} published the most complete catalog of Mira-type stars in the Milky Way. This work significantly expanded the OCVS by discovering over 60,000~Miras toward the Galactic bulge and in the Galactic disk. This study provided pulsation periods, mean magnitudes, and~amplitudes, based on more than two decades of OGLE photometric monitoring. \textcolor{black}{The authors} estimate that the completeness of the catalog is at the level of 96\%. A~set of Mira-type variables from the OCVS is presented in Figure~\ref{fig:iwanek2022}.

\begin{figure}[H]
%\begin{center}
\includegraphics[scale=0.545]{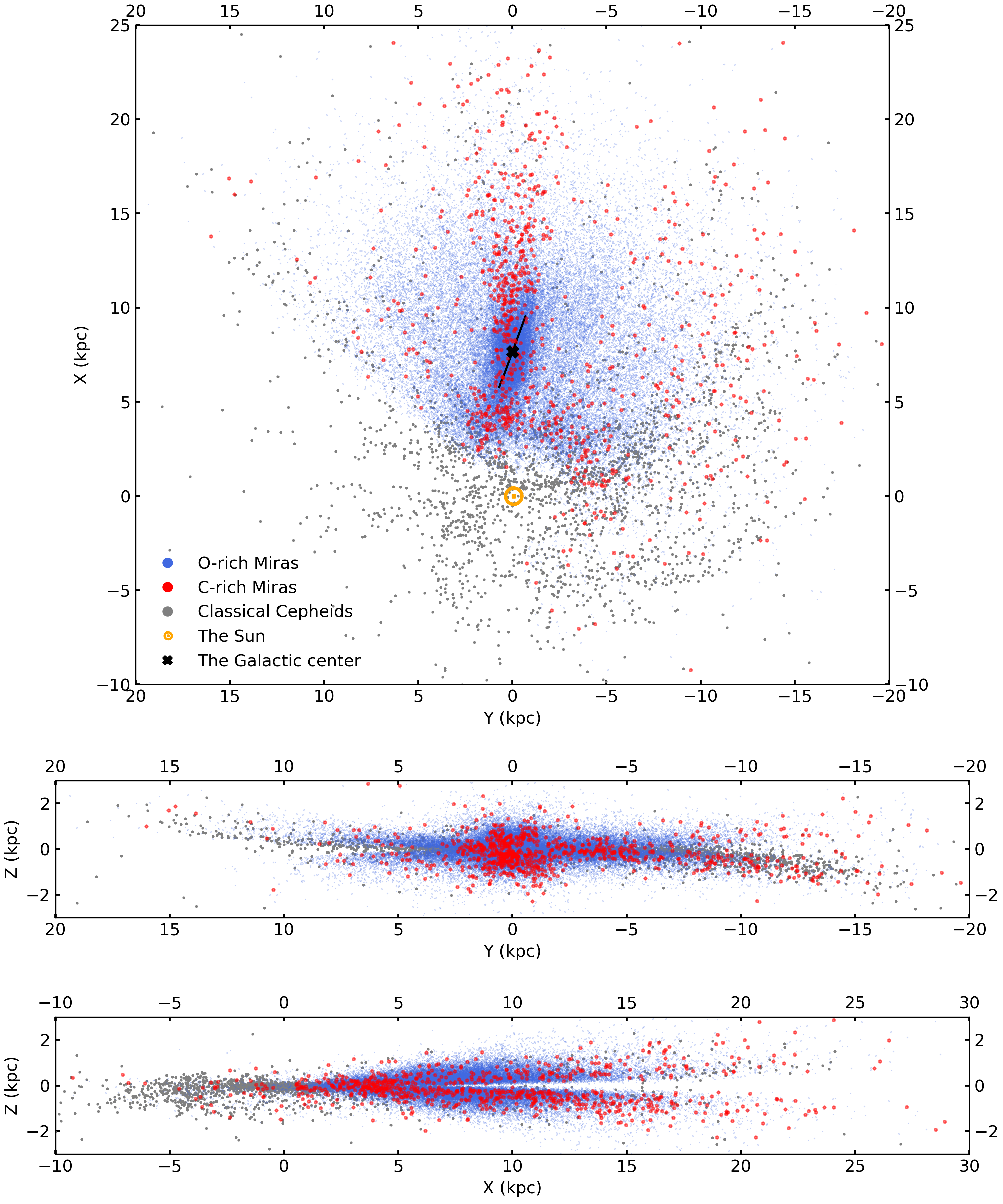}
%\end{center}
\caption{Two-dimensional projection of three-dimensional map of the Milky Way. Blue, red, and~gray points marked O-rich Miras, C-rich Miras, and~classical Cepheids, respectively. Galactic center is marked as black cross, while the black line along the bar shows slope of the bar.  For~more information see \citet{iwanek2023}.}
\label{fig:iwanek2023}
\end{figure}

Based on Miras discovered in the Milky Way, \textcolor{black}{\citet{iwanek2023}} constructed a detailed three-dimensional map of our Galaxy. \textcolor{black}{The authors} analyze the spatial distribution of Mira variables using a~model containing three barred components that include the X-shaped boxy component in the Galactic center, and~an axisymmetric disk \citep{sormani2022}. In~the analysis, \textcolor{black}{they} took into account distance uncertainties by implementing the Bayesian hierarchical inference method. As~a result, \textcolor{black}{the distance to the Galactic center was measured} ($R_0 = 7.66 \pm 0.01$(stat.)$~\pm~0.39$(sys.) kpc), \textcolor{black}{as well as} the inclination of the major axis of the bulge to the Sun-Galactic center line of sight ($\theta = 20.2^\circ \pm 0.6^\circ$(stat.)$~\pm~0.7^\circ$(sys.)), what is consistent with other studies, e.g.,~based on the RR Lyrae stars sample \citep{pietrukowicz2015}. Finally, \textcolor{black}{the authors} showed independent evidence for the X-shaped bulge component. \textcolor{black}{In Figure~\ref{fig:iwanek2023}, the~Milky Way map in three Cartesian projections composed of young (classical Cepheids) and intermediate-age (Mira-type stars) stellar populations is shown.}

\subsection{Blue Large-Amplitude~Pulsators}

Thanks to its vast sky coverage and continuous, long-term monitoring \textcolor{black}{of approximately two billion stars}, the~OGLE project serves not only as a powerful database for studying a well-known type of stellar variability, but~also as a rich source of new and previously unrecognized variable star classes. OGLE has become a true discovery mine, particularly effective in revealing rare and exotic phenomena. One striking example of this is the identification of Blue Large-Amplitude Pulsators (BLAPs).

\begin{figure}[H]
%\begin{center}
\includegraphics[scale=0.25]{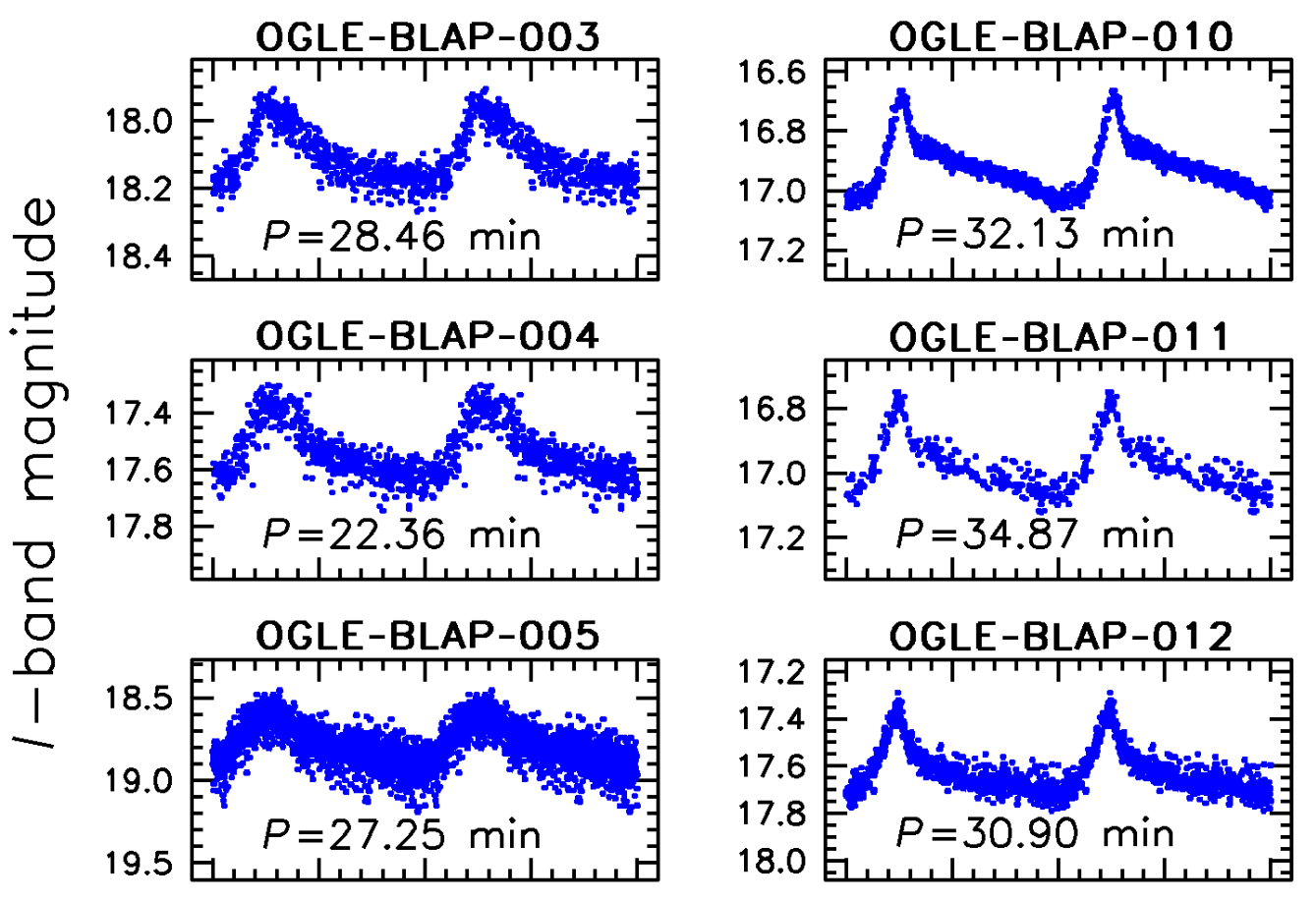}
%\end{center}
\caption{Phase-folded {\textit I}-band light curves of selected BLAPs discovered in the OGLE data. Each light curve is phased with period $P$ indicated in the plot. For~more information see~\mbox{\citet{pietrukowicz2017}}. \textcolor{black}{Reproduced with Springer Nature permission.}}
\label{fig:pietrukowicz2017}
\end{figure}

BLAPs constitute a recently discovered class of variable stars characterized by short pulsation periods---typically from several to several dozen minutes---and large optical amplitudes ($\gtrsim$0.2 mag in the {\textit I}-band, \citep{pietrukowicz2017}). Their phased light curves exhibit a distinctive sawtooth shape, which is shown in Figure~\ref{fig:pietrukowicz2017}. BLAPs are significantly bluer than main-sequence stars observed in the same fields, indicating that they are hot objects, with~a~spectroscopic follow-up confirming effective temperatures around 30,000 K.

The first BLAPs identified by the OGLE have periods in the 20--40 min range, but~subsequent targeted searches in the OGLE-IV data expaded the known range from 7.5 to over 75 min \citep{pietrukowicz2025}. In~the Galactic disk, 25 BLAPs were found---20 of them newly discovered~\citep{borowicz2023a}. A~separate search in the outer Galactic bulge fields revealed 33 additional BLAPs \citep{borowicz2023b}. These discoveries significantly increased the number of known stars of this type to more than~100.

\unskip

The long-term photometric stability of BLAPs, together with observed temperatures and color variations over their pulsation cycles, confirms their pulsation nature. According to pulsation theory, such large-amplitude oscillations over short timescales are expected in evolved, low-mass stars with helium-rich envelopes. However, the~evolutionary path leading to such configurations remains unclear, making BLAPs a key challenge for future stellar evolution~models.

\unskip

\subsection{Long Secondary Periods~Variables}

One of the most puzzling forms of stellar variability observed in evolved stars is the phenomenon of Long Secondary Periods (LSPs). These brightness variations, occurring in roughly one-third of pulsating red giants of the upper Red Giant Branch and Asymptotic Giant Branch, remain the only major type of large-amplitude stellar variability without a~fully established~explanation.

\begin{figure}[H]
%\begin{center}
\includegraphics[scale=0.63]{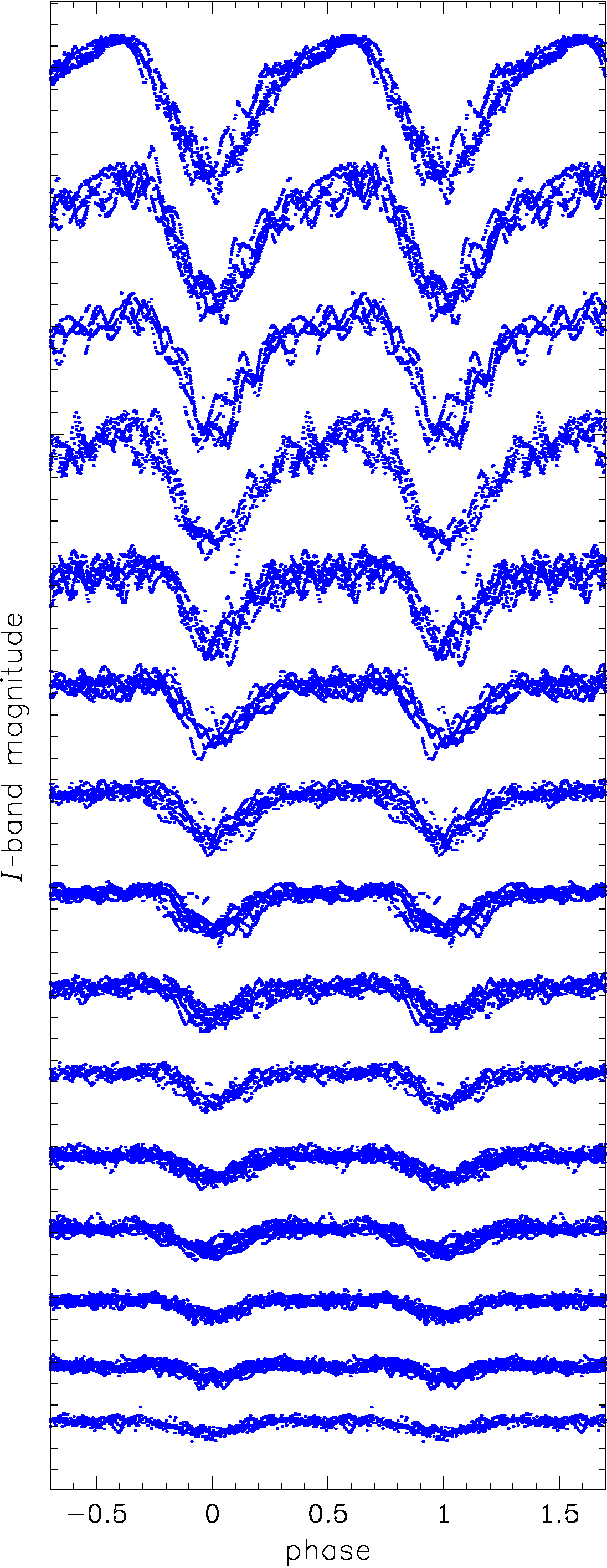}
%\end{center}
\caption{Phased {\textit I}-band light curves of LSPs with different brightness amplitudes of secondary periods discovered in the OGLE data. For~more information see \citet{soszynski2021}. \textcolor{black}{© AAS. Reproduced with permission.}}
\label{fig:soszynski2021}
\end{figure}

The OGLE projects has provided an unparalleled dataset for exploring the LSPs mystery. Over~16,000 well-defined LSP stars have been identified in the OGLE dataset, revealing common light curve characteristics and enabling in-depth statistical and phenomenological studies \citep{soszynski2014}. A~gallery of examples of LSP light curves is presented in Figure~\ref{fig:soszynski2021}.

Multiple evidences now strongly support a binary scenario as the physical origin of LSPs. In~this model, the~red giant is orbited by a low-mass companion---possibly a~former planet that has accreted mass and evolved into a brown dwarf. The~companion is accompanied by a dusty cloud, which obscures the giant one per orbital cycle, causing the observed long-period photometric variations. In~many cases, additional light curve features, such as double-humped modulations resembling ellipsoidal or eclipsing variability, have been detected and share the same periodicity as the LSPs, further supporting the binary~interpretation.

A particular argument comes from the mid-infrared data from the WISE, where roughly half of a carefully selected sample of OGLE LSPs show secondary eclipses---visible only in the infrared---consistent with the obscuring cloud being eclipsed by the red giant~\citep{soszynski2021}.

The puzzle of LSPs has attracted the attention of the broader stellar astrophysics community, culminating in a dedicated European Research Council (ERC) grant: LSP-MIST (101040160), led by Dorota Skowron. This project focuses on disentangling the mechanisms behind LSP variability, \textcolor{black}{using OGLE data together with multiwavelength and spectroscopic observations to shed light on the physical processes potentially driving this phenomenon and their role in late-stage stellar evolution. The~project also aims to combine the observational evidence with theoretical modeling to directly confront the data with stellar evolution theory.}

\subsection{Millinovae}

A recent breakthrough from OGLE's monitoring of the Magellanic Clouds is the discovery of a new class of eruptive variables: millinovae \citep{mroz2024}. These objects bridge the observational gap between classical novae and other accreting binary systems involving white dwarfs, and~challenge the current understanding of X-ray emission from white dwarf~binaries.

These objects show long-duration, symmetric optical outbursts about 1000 times fainter than classical novae, and~exhibit transient supersoft X-ray emission. Unlike classical novae, no signs of mass ejection are detected during their eruptions. This challenges traditional models, where supersoft X-rays are linked to either novae explosions or steady nuclear burning in high-accretion binaries. The~first such case, ASASSN-16oh \citep{maccarone2019}, raised questions about its nature---OGLE has now identified 29 similar systems, suggesting a new class of accreting white-dwarf~binaries.

\subsection{Supernovae}
\textcolor{black}{
Although OGLE was designed primarily for microlensing, it has also become an efficient finder of extragalactic transients. The~OGLE-IV Transient Detection System has operated in near real time in the wide area around the Magellanic Clouds, detecting supernovae and other transients by difference imaging analysis {(DIA,} \cite{alard1998, wozniak2000}) on the standard OGLE frames \citep{wyrzykowski2014}. A~two-year analysis of the Magellanic Bridge ($\sim$65 deg$^2$) yielded \mbox{130 transients,} including 126 supernovae, with~a quantified detection efficiency ($\approx$100\% at peak $I$ $< 18.8$ mag and $\approx$50\% at peak $I$ $\approx 19.7$ mag) \citep{kozlowski2013}. OGLE discoveries also include individual well-studied supernovae, e.g.,~OGLE-2013-SN-079 at $z=0.07$, interpreted as consistent with a helium-shell detonation \citep{inserra2015}. Furthermore, a~dedicated study of \mbox{11 type} II supernovae discovered by OGLE-IV showed that magnitude-limited surveys can preferentially find brighter, often more rapidly evolving type II supernovae~\citep{poznanski2015}.
}

\subsection{Rotating~Variables}

Stellar rotation plays a crucial role in shaping stellar structure and evolution, particularly through its influence on magnetic activity. One of the most prominent manifestations of this activity is rotational modulation caused by starspots---cool or chemically peculiar regions on the stellar surface---and occasional~flares.

\begin{figure}[H]
%\begin{center}
\includegraphics[scale=0.7]{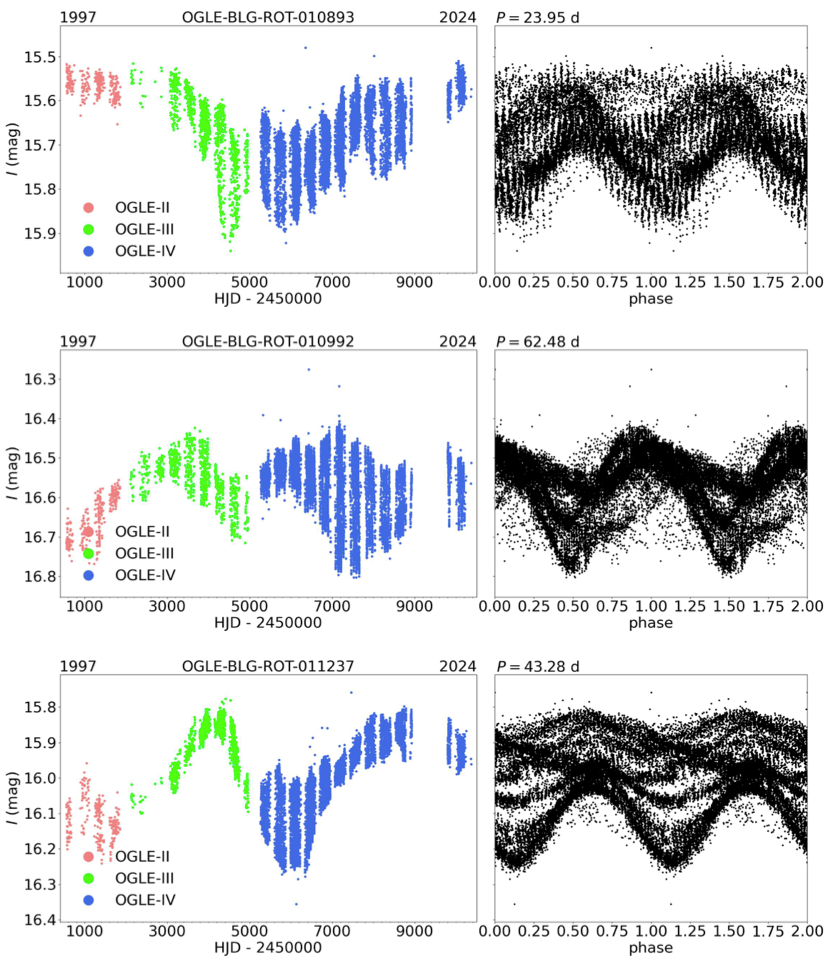}
%\end{center}
\caption{{Three} examples of rotating variables. For~more information see \citet{iwanek2024}.}
\label{fig:iwanek2024}
\end{figure}

In the direction of the Galactic bulge, OGLE has identified and characterized over 18,000 rotating variable stars, including both dwarfs and giants \citep{iwanek2024}. These objects exhibit periodic or quasi-periodic light variations due to the rotation of spotted surfaces, with~amplitudes and periodicities strongly dependent on stellar type and magnetic activity level. Three examples of such variable stars are presented in Figure~\ref{fig:iwanek2024}.

A particular detailed analysis of a subset of 12,660 spotted variables revealed well-defined correlations between brightness, amplitude, and~rotation period, especially among giant stars. Giants show the highest variability amplitudes---up to 0.8 mag in the {\textit I}-band, while rapidly rotating dwarfs ($P \leq 2$ days) tend to have much lower amplitudes ($<$0.2 mag). A~novel dereddening method, tailored for the complex and non-uniform extinction toward the bulge, allowed accurate placement of these stars in the color-magnitude diagram and enabled their classification: 11,812 stars were classified as giants, and~$848$ as~dwarfs \citep{iwanek2019}.

The collection of rotational variables identified by OGLE, with~time-series photometry spanning more than two decades in the {\textit I}- and {\textit V}-band, represents one of the most complete samples investigating stellar magnetic activity and its long-term evolution. This opens the door to studying activity cycles, dynamo processes, and~the relation between stellar structure and rotation in a wide range of stellar~populations.

\subsection{Exoplanets}

The OGLE project has played a pioneering role in the detection of exoplanets, contributing fundamentally to the development and first successful application of both gravitational microlensing and transit~methods.

The idea that binary stars and planetary systems could be detected through the microlensing phenomena, proposed by \citet{mao1991} and \citet{gould1992}, became reality thanks to OGLE's monitoring---starting with the first-ever observed binary microlensing event in 1994 \citep{udalski1994b}. This breakthrough came with the detection of the first microlensing event with a definitive planet detection, in~collaboration with the MOA project \citep{bond2004}. Since then, OGLE has participated in the discovery of hundreds of microlensing planets \citep{cassan2012, zang2025}.

%\unskip

OGLE also made key contributions to the discovery of free-floating planet---planetary-mass objects not bound to any star \citep{mroz2017}. These are identified through extremely short microlensing events, lasting less than a few days. An~example is OGLE-2019-BLG-0551, a~microlensing event characterized by an exceptionally short Einstein timescale of \mbox{$0.381 \pm 0.017$ days}, which strongly suggests a planetary-mass lens and makes it a compelling candidate for free-floating planets \citep{mroz2020b}.

In addition to microlensing, OGLE also pioneered the transit method of exoplanet detection. In~2002, OGLE published the first list of planetary transit candidates \citep{udalski2002}, at~a time when only one transiting planet was known, which was detected first spectroscopically, while transit observations were made afterwards. This campaign led to the discovery of OGLE-TR-56b, the~first exoplanet found by transient observation and confirmed spectroscopically \citep{konacki2003}.

%%%%%%%%%%%%%%%%%%%%%%%%%%%%%%%%%%%%%%%%%%

\section{Summary} \label{sec:summary}

During three decades of operations, the~OGLE survey has changed our understanding of stellar variability and time-domain astrophysics. With~its exceptional cadence, coverage, and~longevity, OGLE has enabled the discovery and precise characterization of more than a~million variable stars in Milky Way and Magellanic Clouds. It has revealed a new classes of variables, deepened our insight into stellar structure and evolution, and~opened new frontiers in exoplanet detection---including the discovery of free-floating planets. OGLE's legacy demonstrates the power of systematic, long-term sky monitoring and continues to provide an irreplaceable foundation for present and future research in~astronomy.

%%%%%%%%%%%%%%%%%%%%%%%%%%%%%%%%%%%%%%%%%%
\vspace{12pt}

\funding{This research was supported by the European Union (ERC, LSP-MIST, 101040160). Views and opinions expressed are, however, those of the authors only and do not necessarily reflect those of the European Union or the European Research Council. Neither the European Union nor the granting authority can be held responsible for them.}

\dataavailability{The entire OGLE Collection of Variable Stars (OCVS) can be accessed via the OGLE website \url{https://ogle.astrouw.edu.pl} (accessed on 11 July 2025) or through the catalog repository: \url{https://www.astrouw.edu.pl/ogle/ogle4/OCVS} (accessed on 11 July 2025).}

\acknowledgments{I thank the anonymous referees for constructive
and valuable feedback that improved this manuscript. I also thank Andrzej Udalski, Igor Soszyński, Paweł Pietrukowicz, and~Przemek Mróz for their comments on the~manuscript.}

\conflictsofinterest{The author declares no conflicts of~interest.}

\begin{adjustwidth}{-\extralength}{0cm}

\setenotez{list-name=Note}

\printendnotes[custom]

\end{adjustwidth}

\begin{adjustwidth}{-\extralength}{0cm}

\reftitle{References}
% ACS format
%\isAPAandChicago{}{%

%}

% If authors have biography, please use the format below
%\section*{Short Biography of Authors}
%\bio
%{\raisebox{-0.35cm}{\includegraphics[width=3.5cm,height=5.3cm,clip,keepaspectratio]{Definitions/author1.pdf}}}
%{\textbf{Firstname Lastname} Biography of first author}
%
%\bio
%{\raisebox{-0.35cm}{\includegraphics[width=3.5cm,height=5.3cm,clip,keepaspectratio]{Definitions/author2.jpg}}}
%{\textbf{Firstname Lastname} Biography of second author}

% For the MDPI journals use author-date citation, please follow the formatting guidelines on http://www.mdpi.com/authors/references
% To cite two works by the same author: \citeauthor{ref-journal-1a} (\citeyear{ref-journal-1a}, \citeyear{ref-journal-1b}). This produces: Whittaker (1967, 1975)
% To cite two works by the same author with specific pages: \citeauthor{ref-journal-3a} (\citeyear{ref-journal-3a}, p. 328; \citeyear{ref-journal-3b}, p.475). This produces: Wong (1999, p. 328; 2000, p. 475)

%%%%%%%%%%%%%%%%%%%%%%%%%%%%%%%%%%%%%%%%%%
%% for journal Sci
%\reviewreports{\\
%Reviewer 1 comments and authors’ response\\
%Reviewer 2 comments and authors’ response\\
%Reviewer 3 comments and authors’ response
%}
%%%%%%%%%%%%%%%%%%%%%%%%%%%%%%%%%%%%%%%%%%
\PublishersNote{}
%\isPreprints{}{% This command is only used for ``preprints''.
\end{adjustwidth}
%} % If the paper is ``preprints'', please uncomment this parenthesis.
\end{document}